\documentclass[3p,10pt,twocolumn]{elsarticle}
\biboptions{sort&compress}
\usepackage{xcolor} 
\usepackage{hyperref}
\hypersetup{colorlinks=true, breaklinks=true, citecolor=blue, linkcolor=blue, menucolor=blue, urlcolor=blue}
\usepackage{siunitx}
\usepackage[normalem]{ulem} 
\DeclareSIUnit \bar{bar}
\DeclareSIUnit \mbar{\milli\bar}
\DeclareSIUnit \ML{ML}
\DeclareSIUnit \atoms{atoms}

\newcommand{\TCNPP}{2H-TCNPP}

\newcommand{\rsCoO}{rs-CoO(100)}
\newcommand{\stmparameter}[2]{\SI{#1}{\V}, \SI{#2}{\nA}}
\newcommand{\rt}{$(\sqrt{58}\times\sqrt{58})\textrm{R}23.2^\circ$}
\newcommand{\figref}[1]{\figurename\ \ref{#1}}
\newlength{\figWidth}
\begin{document}
	\title{Adsorption, self-assembly and self-metalation of tetra-cyanophenyl porphyrins on semiconducting CoO(100) films\tnoteref{t1}}
	\tnotetext[t1]{\href{https://doi.org/10.1016/j.susc.2022.122044}{DOI: 10.1016/j.susc.2022.122044}}
	\author[1]{Maximilian Ammon}
	\author[1]{Andreas Raabgrund}
	\author[1]{M. Alexander Schneider\fnref{fn1}}
	\fntext[fn1]{alexander.schneider@fau.de}
	\address[1]{Solid State Physics, Friedrich-Alexander-Universit\"{a}t Erlangen-N\"{u}rnberg (FAU), Staudtstr. 7, 91058 Erlangen, Germany}

\begin{abstract}
The adsorption properties of free base 5,10,15,20-tetrakis(p-cyanophenyl)porphyrin (\TCNPP) on thin films of rock salt (rs) CoO(100) on Au(111) was studied in ultra-high vacuum (UHV) by a combination of low-temperature scanning tunneling microscopy and spectroscopy (STM/STS) and density functional theory (DFT).   
Films of \rsCoO\ on Au(111) are prepared with excellent quality in a suitable thickness range. 
Particularly, we found that films of only 1\,nm thickness show a semiconducting energy gap of $E_\mathrm{g}=\SI{2.5(2)}{\eV}$.
Upon deposition at 300\,K, \TCNPP\ adsorbs flat-lying and self-assembles in a long-range ordered superstructure that is stable at 80\,K.
The adsorption geometry of the molecules on the surface and within the self-assembly is analyzed by DFT.
We find that the self-assemblies are stabilized by hydrogen bridge bonding via the functional cyano groups.
Our STS data shows molecular states within the fundamental gap of the CoO.
By comparison with the calculated DOS we determine the energetic positions of the frontier orbitals and find that the first three LUMO states \TCNPP\ are located within the band gap, whereas the HOMO is shifted 1\,eV below the CoO conduction band edge.

Upon annealing to 420\,K the molecules change their appearance in STM images and a new prominent electronic state located at the center of the molecule is formed. 
We interpret this changed configuration as Co-TCNPP created by self-metalation on the oxide surface. 

\end{abstract}




	\date{January 16, 2022}
\maketitle

\section{Introduction}

The adsorption of porphyrins on solid surfaces are met with great interest due to their potential in surface functionalization and corresponding applications \cite{Auwaerter2015}.
Most surface-science studies investigated porphyrins on (noble) metal surfaces illustrating the role of functional groups in the creation of self-assembly superstructures and also in (self-)metalation processes \cite{Gottfried2015,Marbach2015,Lepper2018}.
For applications in photo catalysis \cite{Gao2020}, photovoltaics \cite{ORegan1991}, and others it is necessary to decouple the porphyrins from metal substrates either by (ultra-)thin insulating layers \cite{Ramoino2006, Kim2013, Sun2016, Xiang2020} or by using insulating or semiconducting substrates from the beginning \cite{Ashkenasy2000,Lackinger2012,Cristaldi2013}. 
To study the properties of molecules on bulk semiconducting and insulating substrates is met with several problems, most notably that of anchoring the molecules for adsorption studies under clean ultra-high vacuum (UHV) conditions \cite{Rahe2013}.
Therefore studies that address the adsorption geometry on bulk dielectrics in real-space by scanning-probe methods are still of limited numbers today \cite{Maier2008, Amrous2014,Hoff2014,Olszowski2015,Hinaut2015,Lafloer2020}.

An aspect related to the molecule-substrate interaction is whether or not metal atoms from the substrate can be coordinated by adsorbates. 
While porphyrin self-metalation (i.e. the coordination of a substrate atom in the porphyrin core) and the formation of metal-organic intermolecular bonds is quite common on metal surfaces \cite{Gottfried2015}, the fact that porphyrins may be metalated on oxide surfaces like  MgO(100) \cite{Schneider2016, DiFilippo2017}, TiO$_2(110)$ \cite{Wechsler2021}, and CoO(100) and (111) \cite{Wang2020,Waehler2020} is surprising. 
The origin of the metal atoms is ascribed to step edge rather than terrace sites.

Here we investigate the adsorption properties and self-metalation of a functionalized free-base porphyrin on a semiconducting oxide surface by STM. 
We choose rock salt cobalt oxide in the (100) surface orientation (\rsCoO) that by its semiconducting, antiferromagnetic nature is a very potent material for adsorption studies \cite{Elp1991,Kurmaev2008, Xu2016, Schwarz2017, Schwarz2019, Waehler2020}.
We grow films of high crystalline and of thickness larger than \SI{1}{\nm} on Au(111).  
As these films are not in the ultra-thin limit, they may be taken as representative of a bulk oxide surface.

\section{Methods}

\subsection{Experimental}
\begin{figure}
	\centering
	\includegraphics[width=\figWidth]{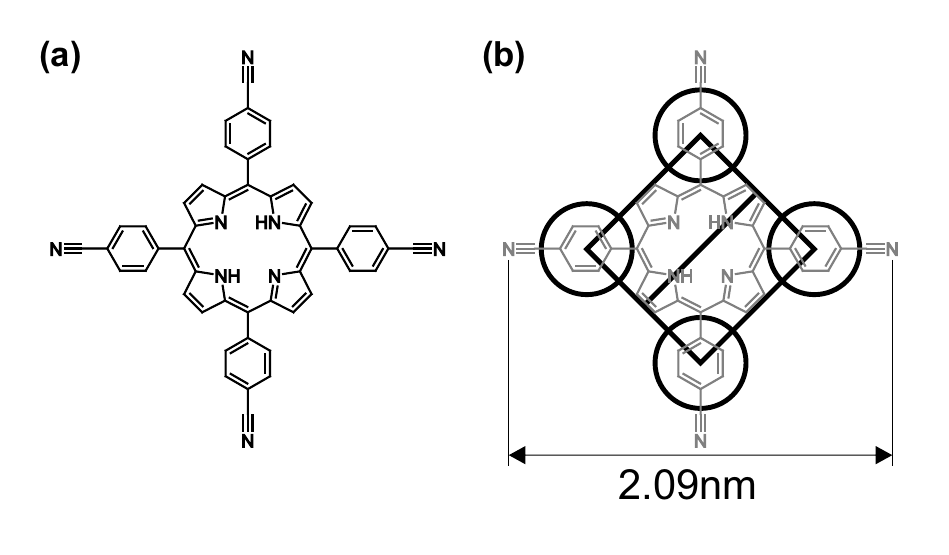}
	\caption{\label{fig:MET_schema}(a) Structure formula of \TCNPP. (b) Schematic representation used to indicate the size and orientation of \TCNPP\ in STM data. Note that the molecular axes are shown as the horizontal and vertical direction respectively, whereas the schematic representation indicates the nodal line in the STM appearance at $45^\circ$ caused by the molecular pyrrole rings.}  
\end{figure}

The experiments were performed in an ultra-high vacuum system, with two separate chambers for sample preparation and STM measurements. 
The base pressure in both chambers was below \SI{1e-10}{\mbar}. For the preparation of the CoO film we used a thin stainless steel tube (“doser”) as oxygen inlet, which was positioned \SI{2}{\cm} in  front of the surface to enhance the local oxygen pressure at the sample by a factor of 50 compared to backfilling the chamber \cite{Ammon2021}. 
All sample temperatures during the preparation were measured with a K-type thermocouple in direct thermal contact to the single crystal.

The Au(111) single crystal was cleaned by repetitive cycles of sputtering with Ne ions (\SI{1.5}{\kV}) and annealing to \SI{800}{\K}. 
Cobalt was evaporated from an e-beam source, the cobalt atoms flux was calibrated with a quartz microbalance and adjusted to \SI{2.1e13}{\atoms\per\s} corresponding to \SI{0.02}{\ML\per\s} where \SI{1}{\ML} of \rsCoO\ contains \SI{1.1e15}{\atoms\per\cm^2} ($a_p = \SI{3.01}{\angstrom}$).
A \rsCoO\ film is formed via temperature-induced restructuring of a wurtzite CoO($000\bar 1$) film as described in our previous publication \cite{Ammon2021} with an optimized procedure as follows.
The initial cobalt oxide was grown reactively under oxygen atmosphere with an $p_\mathrm{O_2} \approx \SI{5e-7}{\mbar}$, while the surface was kept at \SI{320}{\K} rather than at room temperature, since it sped up the sample preparation time without any loss of quality regarding the \rsCoO\ films. 
After the initial reactive growth the sample was annealed in UHV to convert the film from wurtzite cobalt oxide to \rsCoO. 
The exact conversion temperature depends on the film thickness and the growth conditions, so we monitored the conversion by simultaneous LEED measurements. 
For a nominal thickness of \SI{4.4}{\ML} \rsCoO\ an annealing to \SI{800}{\K} was found to be optimal. 

The \TCNPP\ molecules (\figref{fig:MET_schema}) were evaporated from a home-build Knudsen cell evaporator held at a temperature of \SI{650}{\K}.  
The relation between evaporation time and molecular coverage was determined by evaporating onto clean Au(111) to yield approximately half areal coverage.
Additionally, we tested the intactness 
of the molecules by reproducing \TCNPP\ chains on Cu(111) \cite{Lepper2017}, which would fail in case of missing CN end groups.

The homebuilt low-temperature STM was operated at \SI{80}{\K} for all measurements presented here. 
The bias voltage was applied to the sample, while the tip was virtually grounded via the $I/V$-converter. 
The scanning tunneling spectroscopy (STS) measurements were recorded with opened feedback loop using a lock-in technique ($V_\mathrm{rms}=\SI{20}{mV}$, $f=\SI{2.7}{\kHz}$).  
The spectra were normalized to the conductivity $G_\mathrm{sp}= I_\mathrm{sp}/V_\mathrm{sp}$  prior to opening the feedback loop.

\subsection{Computational}

DFT calculations were performed using the the Vienna Ab initio Simulation Package (VASP) \cite{vasp1, vasp2} in the projector augmented wave (PAW) framework \cite{paw}.
The generalized gradient approximation (GGA) of Perdew-Burke-Ernzerhof (PBE) \cite{pbe} for the exchange-correlation functional was employed.
The localized nature of the Co $d$-states was treated by the DFT+U approach of Dudarev with an effective $U_\mathrm{eff}= (U-J)=\SI{4}{\eV}$ \cite{ldaU} and $U_\mathrm{eff}=\SI{2}{\eV}$ for the Co atom in the Co-TCNPP molecule \cite{Leung2010}.
To account for van der Waals forces we used the Tkatchenko-Scheffler (TS) scheme \cite{TS_2009}.
The substrate was modeled by a four CoO layers thick, (100) oriented slab with \SI{20}{\angstrom} vacuum between its repetitions along the [100] direction.
\TCNPP\ or Co-TCNPP was set on one of the slab surfaces, all molecular atoms and the top two CoO layers were allowed to relax, the two bottom CoO layers were kept at the bulk structure (with the appropriate DFT lattice parameter of $a_\mathrm{p}= \SI{2.97}{\angstrom}$ as obtained from a corresponding bulk CoO calculation).
The relaxation was done in a non-spin-polarized, $\Gamma-$point only calculation until forces were below \SI{0.1}{\eV \per \angstrom }, the lateral dimensions of the slab were chosen according to the experimental unit cell (see below).
To model electronic properties we used a four layer ($8 \times8$) \rsCoO\ slab, carried out the relaxation as before and performed a spin-polarized DFT+U+TS self-consistent calculation converged to $\Delta\,E / E = 5\cdot 10^{-6}$ taking into account the antiferromagnetic order of CoO.
We ensured that the DFT+U approach ended in the lowest energy configuration for bulk CoO by choosing an arbitrarily high $U_\mathrm{eff}$ first and then reducing in steps until  $U_\mathrm{eff}=\SI{4}{\eV}$.
This procedure was repeated for the CoO slab with the adsorbed molecule. 

\section{Results and Discussion}

\subsection{Characterization of the substrate}
\begin{figure}
\centering
\includegraphics[width=\figWidth]{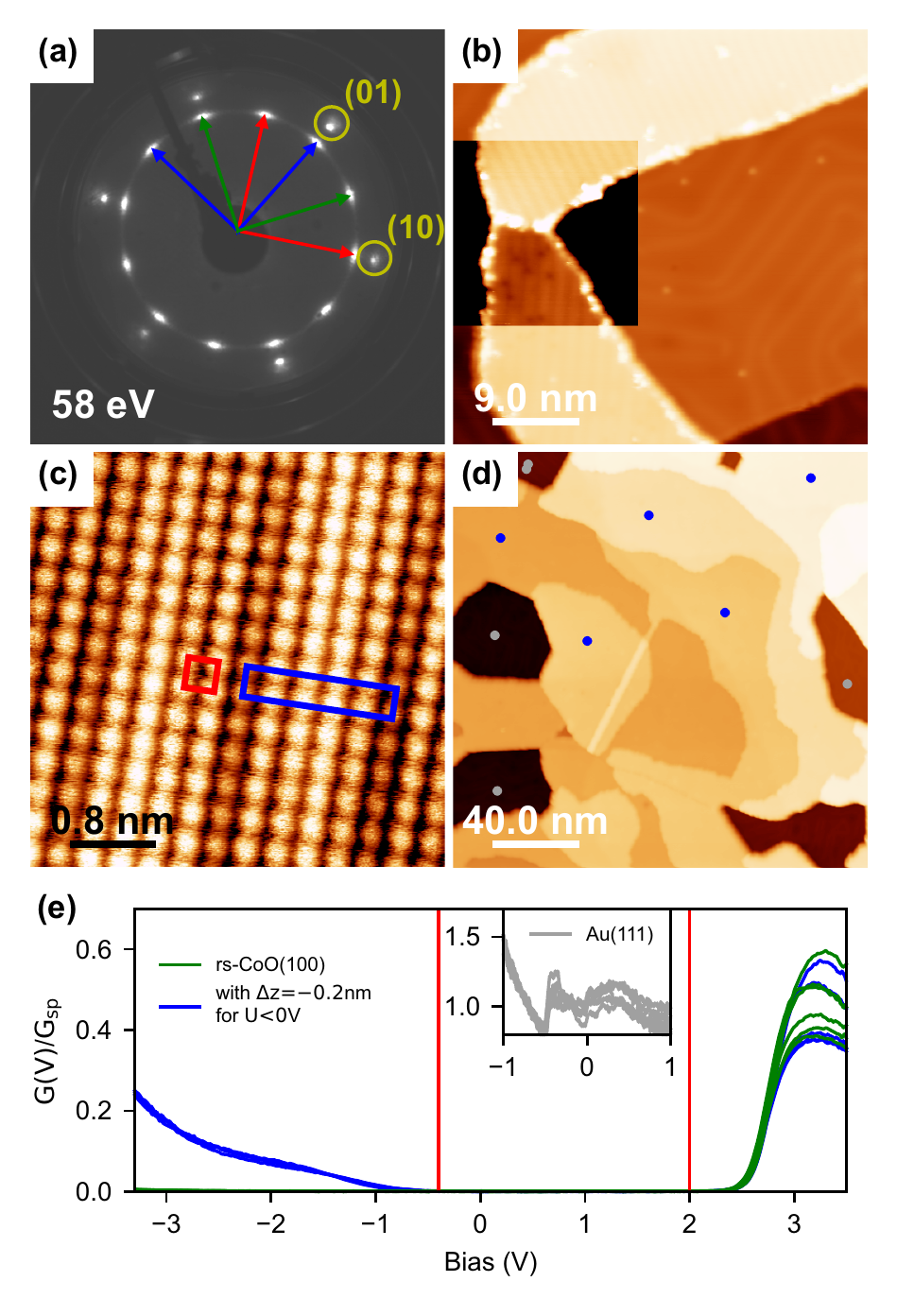}
\caption{\label{fig:STM_rsCoO100} 
		(a) LEED pattern of a nominally \SI{1.9}{\ML} thick \rsCoO\ film. The  Au(111) diffraction spots (yellow circles) are still visible. The reciprocal unit cells of the three rotational domains of the \rsCoO\ film  are marked with red, green and blue arrows. 
	(b) The STM topography shows a flat film surface with a weak moir\'e pattern whose orientation and periodicity depends on the grain orientation to the surface.  
	(c) Atomically resolved \rsCoO\ surface showing a moir\'e that is locally reminiscent of a ``$ 5\times 1$'' with height variations of a few picometer. (red: primitive unit cell, blue: local moir\'e cell). 
	(d) Overview STM image of a nominally \SI{4.4}{\ML} thick \rsCoO\ film. The markers show the positions of the STS data depicted in (e). 
	(e) Normalized differential conductivity curves showing a bandgap from \SI{-0.4}{\eV} to \SI{2.0}{\eV} (red vertical lines). 
	The tip was approached  \SI{2}{\angstrom} closer to the surface in the negative bias region enhancing the signal by a factor of $\approx 100$. Insert: spectra taken with the same tip on clean Au(111). STM parameters: (b) \stmparameter{1.0}{0.10}; (c) \stmparameter{0.3}{0.15}; (d) \stmparameter{3.0}{0.10} (e) prior to opening the FB loop defining $G_\mathrm{sp}$: \stmparameter{-3.5}{0.05} on \rsCoO\ and \stmparameter{-1.0}{0.4} on Au(111).}
\end{figure}

The preparation recipe for the \rsCoO\ films may be tuned to produce discontinuous films with large \rsCoO\ patches and clean Au(111) areas side by side. 
For the present study such substrate preparations were used to allow for a direct comparison between adsorption of \TCNPP\ on Au(111) and \rsCoO\ and for the possibility of STM tip characterization and preparation for STS measurements.
Therefore, films with a nominal thickness of \SIrange{0.4}{4.4}{\ML} were studied in detail in the following. 
The nominal thickness is a measure of the amount of cobalt in the CoO film expressed by the thickness of a hypothetical, homogeneous and closed film. 
Preparations with a nominal thickness of \SIrange{1.0}{4.4}{\ML} led  to open \rsCoO\ films, while e.g. for \SI{5.7}{\ML} completely closed \rsCoO\ film was produced.
\figref{fig:STM_rsCoO100}(a) depicts a LEED measurement of a nominally \SI{1.9}{\ML} thick oxide film, \figref{fig:STM_rsCoO100}(b) and (c) the corresponding STM measurement. 
In contrast, \figref{fig:STM_rsCoO100}(d) shows an STM image of a nominally \SI{4.4}{\ML} thick film. 

The ring of 12 diffraction spots in (a) corresponds to three rotational domains, as expected for the epitaxial growth of a rs(100) film on fcc(111).
The diffraction spots of Au(111), including the moir\'e spots of the herringbone reconstruction, are still visible and indicate intact and clean Au(111) surface areas alongside the \rsCoO\ islands. 
The comparison of the \rsCoO\ and Au(111) spot positions allows to measure the lattice constant of the oxide film with great accuracy. 
We evaluated 18 independently prepared open oxide films with a nominal thickness from \SI{0.4}{\ML} to \SI{4.4}{\ML} and find a mean lattice constant of $a_p = \SI{2.92(1)}{\angstrom}$ with no systematic dependence on the nominal film thickness in that range. 
We note, that thicker and closed films approach the bulk surface lattice parameter of $a_p = \SI{3.01(1)}{\angstrom}$ \cite{Ammon2021}. 
STM measurements of the \SI{1.9}{\ML} films show that roughly \SIrange{30}{50}{\percent} of the surface is covered with oxide (see \figref{fig:STM_rsCoO100}(b)). 
The surface of the oxide film is atomically flat but shows a weak one-dimensional moir\'e pattern with a height variation below \SI{10}{\pm} (see area of enhanced contrast in \figref{fig:STM_rsCoO100}(b)). 
The moir\'e orientation and periodicity may differ from patch to patch which indicates some variations in the orientation of the oxide islands with respect to the substrate, in line with the radial smearing of the diffraction spots in the LEED pattern (see \figref{fig:STM_rsCoO100}(a)). 
In atomically resolved STM images the moir\'e appears most of the time as quasi ``$ 5\times 1$'' ( \figref{fig:STM_rsCoO100}(c)). 
However, the fine details of the STM images reveal that this is only a local section of a much larger (maybe even incommensurate) unit cell ($\SI{2.92}{\angstrom} / \SI{2.88}{\angstrom} \approx 73/72$) in agreement with the LEED pattern.
Therefore the oxide overlayer accommodates itself on the Au(111) similar to what has been found for the case of a single bilayer CoO(111) on Ir(100) \cite{Troeppner12}. 

\figref{fig:STM_rsCoO100}(d) shows a nominally \SI{4.4}{\ML} thick \rsCoO\ film, that covers \SI{85(10)}{\percent} of the surface.
The oxide film exhibits mono-atomic steps and terraces extending over several tens of nanometers, a situation similar to an ideal \rsCoO\ surface.
Since \rsCoO\ is a semiconductor, the apparent height is bias dependent.
We therefore used bias voltages $V_\textrm{bias}>\SI{2}{\V}$ that allows tunneling into the conduction band and which should give a apparent height comparable to the geometric height.
From images like \figref{fig:STM_rsCoO100}(d) the measured apparent height of the oxide surface with respect to the clean Au(111) is \SI{1.2(3)}{\nm} (\SI{5(1)}{\ML}). 
The areal coverage and deposited amount of Co is used to calculate the average film thickness, which results in \SI{1.4(2)}{\nm} (\SI{5.2(6)}{\ML}) and compares well with the measured apparent height.

\figref{fig:STM_rsCoO100}(e) shows a series of STS measurements on large CoO terraces. 
The measurement positions are marked in \figref{fig:STM_rsCoO100} (d). 
The tip quality was tracked by measuring reference spectra on Au(111) showing the fingerprint of the surface state (see \figref{fig:STM_rsCoO100}(e), inset). 
The \rsCoO\ shows a semiconducting behavior with a wide bandgap, where $dI/dV(V)$ and $I(V)$ are zero. 
We observe a much stronger rise of the differential conductance at the conduction than at the valence band edge. 
To obtain suitable sensitivity the STS data of \figref{fig:STM_rsCoO100}(e) was taken for the negative and positive bias region separately whereby the tip was approached \SI{0.2}{\nm} closer to the surface at negative bias. 
From this kind of data we extracted more accurate values for the band onset by fitting the signal with exponential functions in the onset regions. 
Details of the fitting procedure can be found in Ref.\,\cite{Ammon2021}. 
The valence band onset is at $E_\mathrm{V}=\SI{-0.4(1)}{\eV}$, the conduction band onset at $E_\mathrm{C}=\SI{2.0(2)}{\eV}$ which leads to a measured bandgap of $E_\mathrm{g}=\SI{2.5(2)}{\eV}$.
This compares well to $E_\mathrm{g}=\SI{2.5(3)}{\eV}$ found by a combination of photoemission and bremsstrahlung isochromat spectroscopy or optical spectroscopy  \cite{Elp1991,Kurmaev2008,Baraik2020}. An asymmetric position of $E_\textrm{F}$ within the semiconducting gap was also observed by photoemission spectroscopy for w-CoO(100) films on Au \cite{Ammon2021} and CoO(100) films on Ir(100) \cite{Otto2016} and therefore is presumably no artifact of the STS measurement.  
 
\subsection{\TCNPP\ on \rsCoO: self-assembly}

\begin{figure}[ht]
	\includegraphics[width=\figWidth]{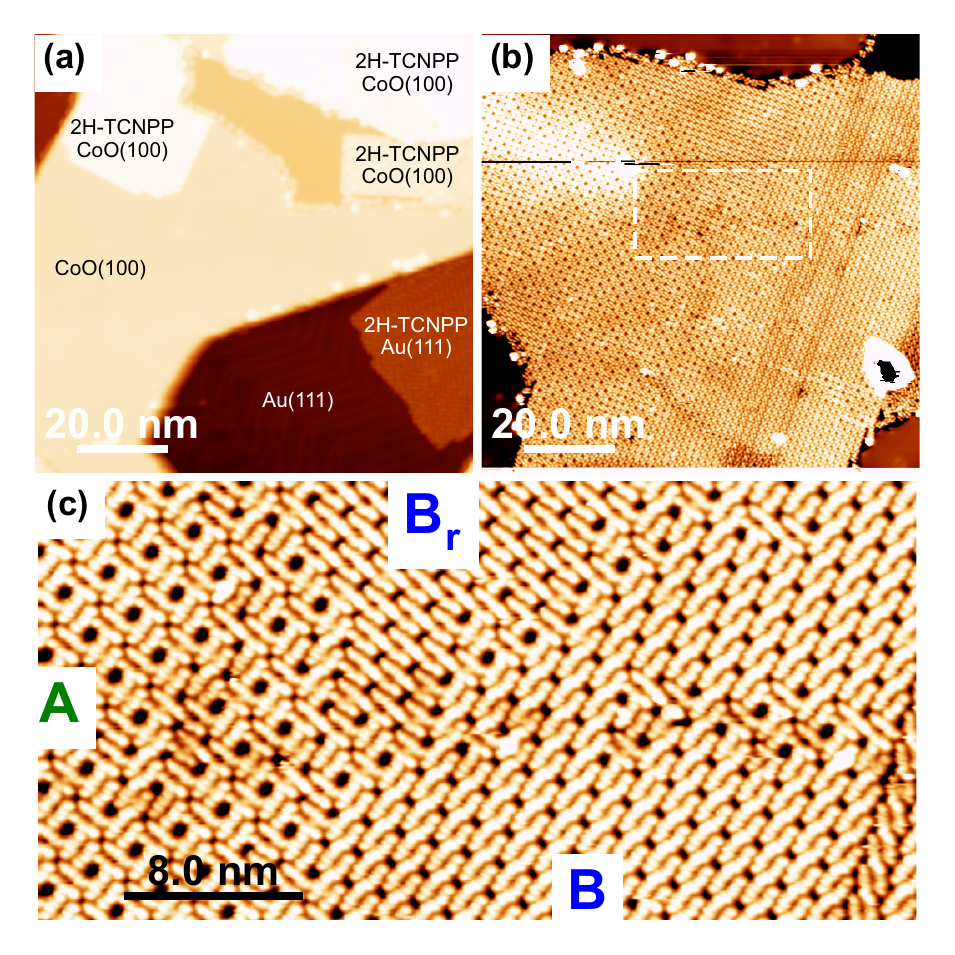}
	\caption{\label{fig:STM_RTDep}STM images taken at 80\,K of submonolayer \TCNPP\ deposited at \SI{300}{\K} on  nominally \SI{4.4}{\ML} thick \rsCoO\ on Au(111):
	(a) Clean Au(111), clean CoO(100) and compact islands of \TCNPP\ on both areas are found. 
	(b) Large scale self-assembly of \TCNPP\ molecules on the \rsCoO\ film. The area marked by the dashed rectangle is depicted in (c). (c) The self-assembly shows two distinct binding motifs, marked with $A$ and $B$. 
	Molecules in motif $A$ are alternately rotated by \SI{90}{\degree}, while in motif B all molecules have the same orientation. 
	Consequently the motif $A$ shows a $C_4$ symmetry, while $B$ is only $C_2$ symmetric and allows for domains $B_r$  rotated by $90^\circ$. 
		STM parameters: (a) \stmparameter{3.5}{0.05}; (b) \stmparameter{3.0}{0.05}; (c) \stmparameter{2.5}{0.05}.}
\end{figure}

\figref{fig:STM_RTDep} shows the result of the deposition of \TCNPP\ at \SI{300}{\K} with submonolayer coverage on a nominally \SI{4.4}{\ML} thick \rsCoO\ film. 
After cooling the sample to 80\,K the molecules are found on the oxide as well as on the free metal. 
The great majority assembles into stable islands (\figref{fig:STM_RTDep}(b) and (c)). 
On the \rsCoO\ surface, isolated \TCNPP\ molecules are only found attached to step edges. 
This indicates that the molecules are mobile at RT or even below and that there are intermolecular forces stabilizing the assemblies.
The apparent height of isolated molecules measured by STM at a bias voltage of 3.5\,V yields \SI{0.25(4)}{nm} which can be taken as proof that the molecules are flat lying similar to findings of other porphyrins on oxide surfaces \cite{Schneider2016, Egger2021, Wang2014a, Wolfram2022}. 

\begin{figure}[t!]
	\centering
	\includegraphics[width=\figWidth]{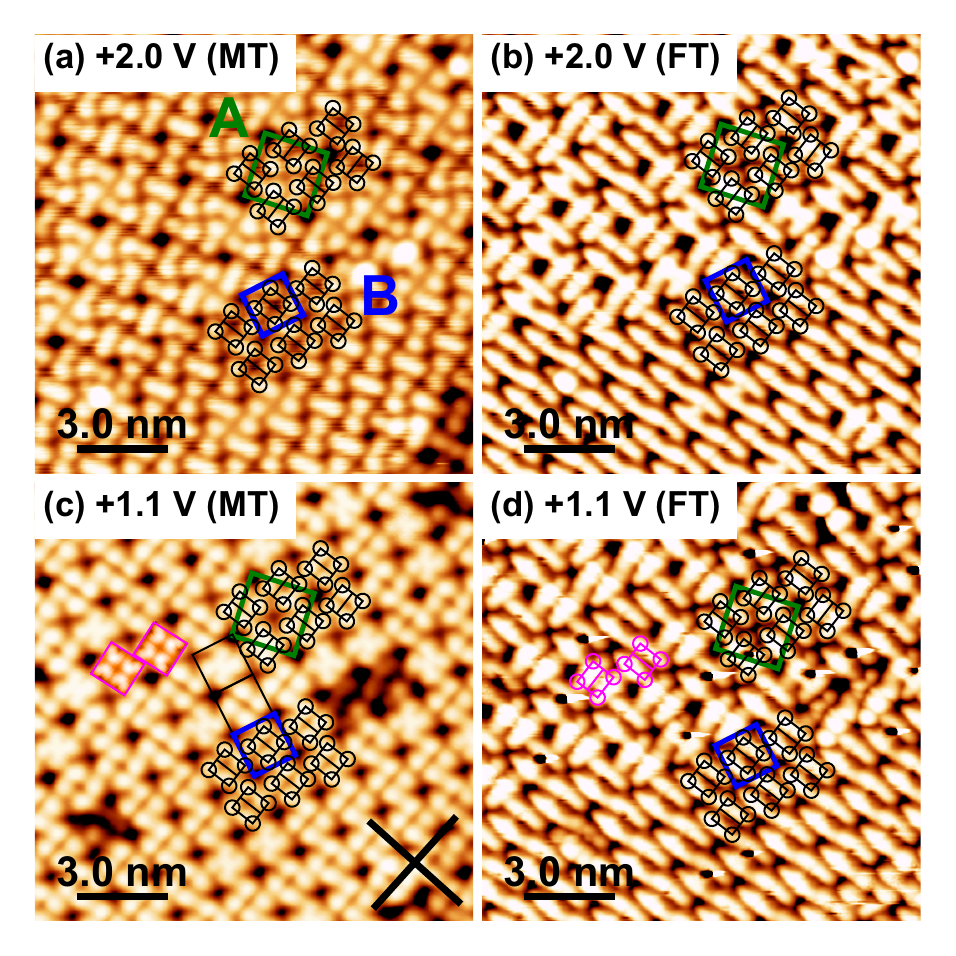}
	\caption{\label{fig:STM_RTSA}Self-assembly of \TCNPP\ on CoO(100) after deposition on 300\,K. The 2D self-assemly of \TCNPP\ measured with a ``metallic'' and a ``functionalized'' tip (MT, FT) at two bias voltages. The position of identical molecules is indicated schematically in all four panels. (a) The STM contrast at \SI{2.0}{\V} with an MT tip mainly shows the four phenyl groups of the molecule whereas in (b) with an FT tip the molecular axis becomes more pronounced and the relative orientation between molecules in the two binding motifs A and B  can be clearly distinguished. The cell of motif A (green cell) is a $(\sqrt{2}\times\sqrt{2})\textrm{R}45^\circ$ with respect to B (blue cell). (c) The measurement at \SI{1.1}{\V} with an MT tip is dominated by the molecular core as judged from the distance of the four maxima making up the molecule that are closer together than in (a). 
		The majority of the molecules is imaged with a height minimum in the molecular center. 
		The tiles with pink border are spin-averaged DFT+U STM image simulations corresponding to the molecules marked in panel (d). Measurement and simulation show a close to fourfold symmetry of the molecules at that bias. (Details see text.)  
		The black cross in the lower right corner defines the [011] and [0$\bar 1$1] directions of the \rsCoO\ substrate.
		(d) The ``functional'' imaging contrast is less bias dependent, hence the differences between (b) and (d) are minor.
		STM-Parameter: (a, b) \stmparameter{2.0}{0.05}; (c, d) \stmparameter{1.1}{0.05}.}
\end{figure}

However, the exact appearance of the molecule on the semiconducting substrate depends sensitively on the properties of the STM tip. 
This is demonstrated in \figref{fig:STM_RTSA} where we compare images taken at exactly the same position, a domain boundary of both motifs, at different STM bias and tip conditions. 
The images of the left-hand column (a) and (c) represent the most robust tip state. 
Here the appearance depends very drastically on voltage as might be expected for a molecule on a semiconducting substrate in the spirit of the Tersoff-Hamann theory of STM \cite{Tersoff-Hamann}.
Hence we term this tip as ``metallic'' (MT).
The images of the right-hand column do not show a particularly strong bias dependence. 
In contrast, fine intramolecular details (like nodal structures) can be detected that are not visible in the left-hand MT images.
This observation together with the reduced stability of this imaging mode leads us to assign such images to ``functionalized'' tips (FT).

In the gas phase, the \TCNPP\ molecule has $\textrm{C}_2$ symmetry due to the aminic (-NH-) and iminic (=N-) nitrogen atoms in the porphyrin macrocycle. 
On the surface, this symmetry manifests itself by a dark ``central line'' in the molecular STM images. 
The appearance of the molecule is very similar to that observed on metal surfaces, where it is attributed to a saddle-shape conformation \cite{WeberBargioni2008, Brede2009, Lepper2017b, Adhikari2020}. 
Hence the difference between assembly patterns A and B are simply different neighboring tautomers of \TCNPP. 
These manifest themselves in the direction of the ``central line'' that is well defined in the FT contrast in \figref{fig:STM_RTSA}. We therefore use the contrast of the FT images to assign the correct orientation of the molecules. STM image simulations employing the spin-resolved DFT+U calculations shown as tiles in  \figref{fig:STM_RTSA}(c) suggest that the orientation may only with great difficulties be inferred from the MT contrast images.
However, this arrangement of the tautomers has consequences for the relative orientation of the cyanophenyl groups (see Section \ref{sec:DFT}). 
The mechanism of the functionalization of the STM tip is unclear, the pick-up of a simple rest-gas molecule (H$_2$, CO, O$_2$) seems likely \cite{Krej2017, Weiss2010}.

The square unit cells of the assemblies are indicated in \figref{fig:STM_RTSA}(a-d). 
Since there are uncovered \rsCoO\ areas we may determine the underlying crystallographic directions of the substrate (\figref{fig:STM_RTSA}(c)). 
With respect to the \rsCoO\ lattice we find that the cell is rotated by $\alpha = \SI{20(4)}{\degree}$ and its lattice parameter is $a=\SI{2.2(1)}{\nm}$. 
The closest commensurate superstructure of the self-assembly of motif A is a \rt-cell with $a=\sqrt{58} \times \SI{0.292}{\nm}\approx\SI{2.2}{\nm}$ that contains two \TCNPP\ molecules (\figref{fig:STM_RTSA}(c)). 
The superstructure B is only half the size of that of A but the adsorption site of the molecules in both structures is the same, as proven by the continuation of the B unit cell into the area of motif A by the black squares in \figref{fig:STM_RTSA}(c).
The appearance of this assembly is identical to that of \TCNPP\ on two-bilayer rs-CoO(111) on Ir(100) \cite{Xiang2020}.
Upon reinspection of that data we discovered that the lateral calibration of the STM was wrong by approximately 10\% and therefore the size of unit cell was determined too large. 
As a consequence, the dipolar interaction between parallel cyano end groups proposed in \cite{Xiang2020} seems rather unlikely. See section \ref{sec:DFT} for a discussion of intermolecular forces stabilizing the self-assemblies.

\subsection{\TCNPP\ on \rsCoO: electronic properties}

\begin{figure}[t!]
	\centering
	\includegraphics[width=\figWidth]{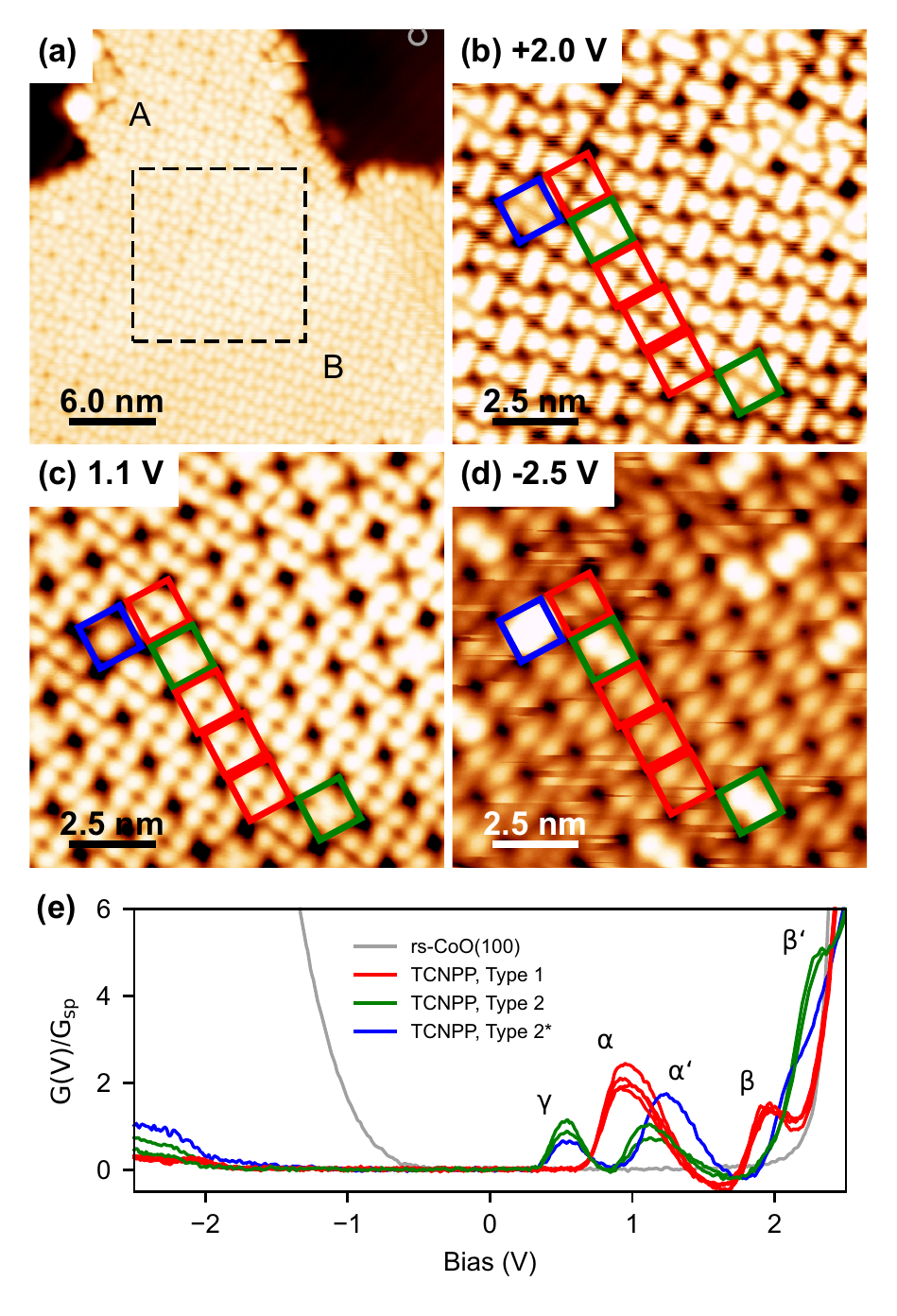}	
	\caption{\label{fig:STM_RTSTS}(a) Overview STM image of a \TCNPP\ island with both binding motifs A and B. The dashed lines mark the area shown in panels (b-d). 
	Three distinct molecular conformations can be discriminated from their appearance in (b)--(d): the majority configuration (red, type 1), a second (green, type 2) that is imaged as a projecting ``cross'' in (c), and a third configuration (blue, type 2*) that looks similar to the second configuration but is imaged as a spherical symmetric blob in (c) and (d). (d) The bone-shaped HOMO orbital shows the molecular orientation of the molecules. 
	Normalized differential conductance STS spectra in (e) are taken with the tip positioned on the centers of the marked molecules in (b)--(d). 
	For comparison, the spectrum on the bare \rsCoO\ film taken with the same tip is shown also.
STM parameters: (a) \stmparameter{2.0}{0.05}; (b) \stmparameter{2.0}{0.05}; (c) \stmparameter{1.1}{0.05}; (d) \stmparameter{-2.5}{0.05}; (e) prior to opening the feed-back loop $G_\mathrm{sp}:$ \stmparameter{2.5}{0.4}.}
\end{figure}

Within the assemblies we find three distinct molecular configurations after deposition of \TCNPP\ on \rsCoO\ at 300\,K.
In \figref{fig:STM_RTSTS} we analyze their electronic properties. 
Not surprisingly we do not find any difference between molecules assembled in motif A or B.

The majority configuration (marked in red, \figref{fig:STM_RTSTS} (b)-(d)) shows the MT contrast discussed in the previous section and a ``double cigar'' contrast at large negative voltages.
Its spectroscopic signature in the unoccupied states region ($V_{bias} > \SI{0}{V}$) consists of prominent peaks $\alpha$ at $V_{bias}=\SI{0.9(1)}{\V}$ and $\beta$ at $V_\mathrm{bias}=\SI{2.0(1)}{\V}$ (\figref{fig:STM_RTSTS}(e)).
The cross-like (type 2) molecules (marked in green, \figref{fig:STM_RTSTS} (b)-(d)) show an additional  peak $\gamma$ at $V_\mathrm{bias}=\SI{0.6(1)}{\V}$ and the other two peaks shifted to higher energies.  
The previous $\alpha$-peak (now $\alpha^\prime$) is centered at $V_\mathrm{bias}=\SI{1.1(1)}{\V}$ and $\beta^\prime$ is shifted to higher energies and mingles with the conduction band onset above \SI{2}{V}. 
Most striking the type 2 molecules are imaged clearly differently for negative voltages and can be discriminated against a third configuration (type 2*, marked in blue, \figref{fig:STM_RTSTS} (b)-(d)) that is imaged similarly at $V_\mathrm{bias}=+\SI{2.0}{V}$ but has its  $\alpha^\prime$ feature at \SI{1.3(1)}{\V}. 
Type 2/2* molecules have a dome-like appearance at voltages below $\alpha^\prime$, since only a state $\gamma$ located at the core center contributes to the conductance. 
All molecular configurations show a very shallow increase of the differential conductivity for negative bias. 
Type 1 shows the weakest signal with an onset for the occupied orbitals at \SI{-1.7(3)}{\V}, the types 2 and especially 2* show stronger signals, so that the onset is clearer defined and results in \SI{-1.8(1)}{\V} for both. 
For all molecular configurations we find a substantial suppression of the STS signal from the valence band edge (gray curve in \figref{fig:STM_RTSTS}(e)).
This indicates the absence of molecular states in that energy region. As a consequence there is a very low tunneling transmission from valence band edge states due to the much larger distance between tip and CoO surface when probing a molecule.

\subsection{\TCNPP\ on \rsCoO: thermally induced self-metalation}

The existence of different molecular configurations on the surface calls for an investigation of its reason. 
Possibilities that can be derived from literature are the occurrence of distinctly different porphyrin geometries, e.g. saddle-shape vs. dome-like \cite{WeberBargioni2008, Brede2009, Ditze2014, Stark2014}, or molecular reactions like the deprotonation \cite{Auwaerter2012} or metalation of molecules \cite{Marbach2015, Diller2016}. 

\begin{figure}[t]
	\centering
	\includegraphics[width=\figWidth]{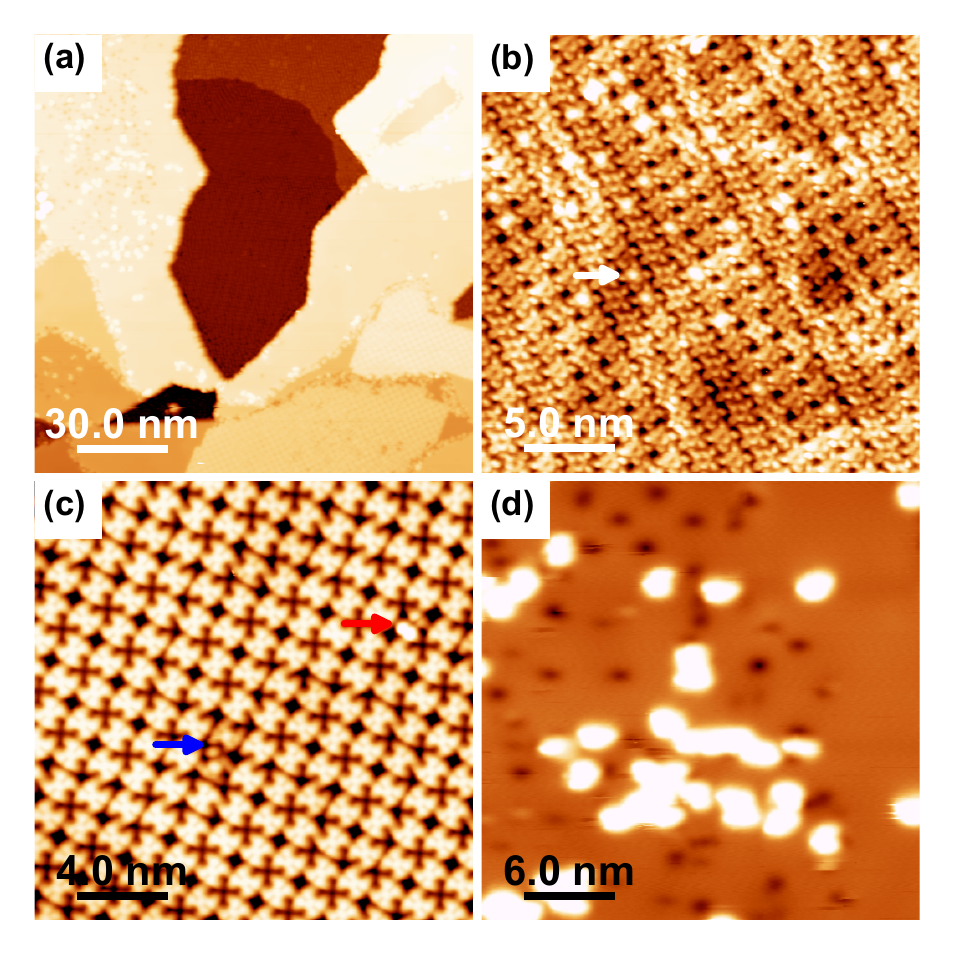}	
	\caption{\label{fig:STM_An} (a) Overview after annealing to \SI{420}{\K}. 
		(b) \TCNPP\ on the Au(111) area. The majority  of Au(111) areas are completely filled with \TCNPP\ molecules which suggests diffusion from the oxide to the metal substrate during annealing. 
		Only few molecules have a filled core (e.g. white arrow) that are interpreted as Co-TCNPP. 
		(c) The self-assembly \TCNPP\ on \rsCoO\ got more homogeneous, defects are healed out. 
		Now most molecules are of type 2 (x-shaped, filled center). 
		These are interpreted as Co-TCNPP molecules created by self-metalation on \rsCoO.
		The red arrow marks some exceptional type 1, the blue arrow marks two type 2* molecules.
		(d) Defect rich areas can be found on the \rsCoO\ substrate. Only single molecules are adsorbed here. 
		Such defects were not observed previously on the \rsCoO\ substrate and could hence be the sites where part of the Co atoms for metalation originated. STM parameters: (a) \stmparameter{3.0}{0.05}; (b) \stmparameter{1.0}{0.05}; (c) \stmparameter{1.35}{0.05}; (d) \stmparameter{3.5}{0.05}.}
\end{figure}

From the appearance of the type 2 molecules with their filled center (at least visible at certain voltages), the metalation reaction is a possibility supported by the findings on metals \cite{Marbach2015, Diller2016, Lepper2018}. 
The rate of a metalation reaction is expected to depend on temperature \cite{Marbach2014}.
Metalation of porphyrins on cobalt oxide has also been observed before \cite{Wechsler19}.
It is known (also established by our own experiments) that the \TCNPP\ do not self-metalate on Au(111). 
Hence, the only possible candidate for the outcome of a metalation reaction is Co-TCNPP for which Co atoms have to be provided by the oxide.
In \figref{fig:STM_An}(a) we show the situation after annealing to \SI{420}{\K} for \SI{8}{\min}. 
We observe that the remaining bare Au(111) surfaces are completely covered with molecules in an ordered self-assembly (\figref{fig:STM_An}(b)). 
Only some molecules within the assemblies show spectroscopic and topographic signatures of a metalated porphyrin core, like the commonly observed protrusion in the center \cite{Marbach2015, Diller2016}, which arises from an additional state close to $E_\textrm{F}$ \cite{Papageorgiou2013, Urgel2015, Diller2016, Bischoff2018}. 
The ratio between empty and filled molecules on Au(111) varies from place to place, but metalated molecules were the minority species always.
In contrast, our experiments with \TCNPP\ on Au(111) indicate that with an appropriate (small) amount of Co atoms \TCNPP\ molecules are metalated very efficiently at 420\,K.
Therefore, we conclude that the CoO does not produce a Co adatom gas on the bare Au(111) areas.

Conversely, most molecules remaining on the \rsCoO\ surface show now the appearance of type 2 molecules. 
These molecules assemble predominantly into well ordered structures of motif A (\figref{fig:STM_An}(c)). 
On some but not on all bare \rsCoO\ surfaces we observe now defects that were not present after the preparation and deposition of \TCNPP\ at 300\,K (\figref{fig:STM_An}(d)). 
Also some isolated molecules are found in these defective areas that are of an unknown configuration and were not observed before. 

\begin{figure}[t!]
	\centering
	\includegraphics[width=\figWidth]{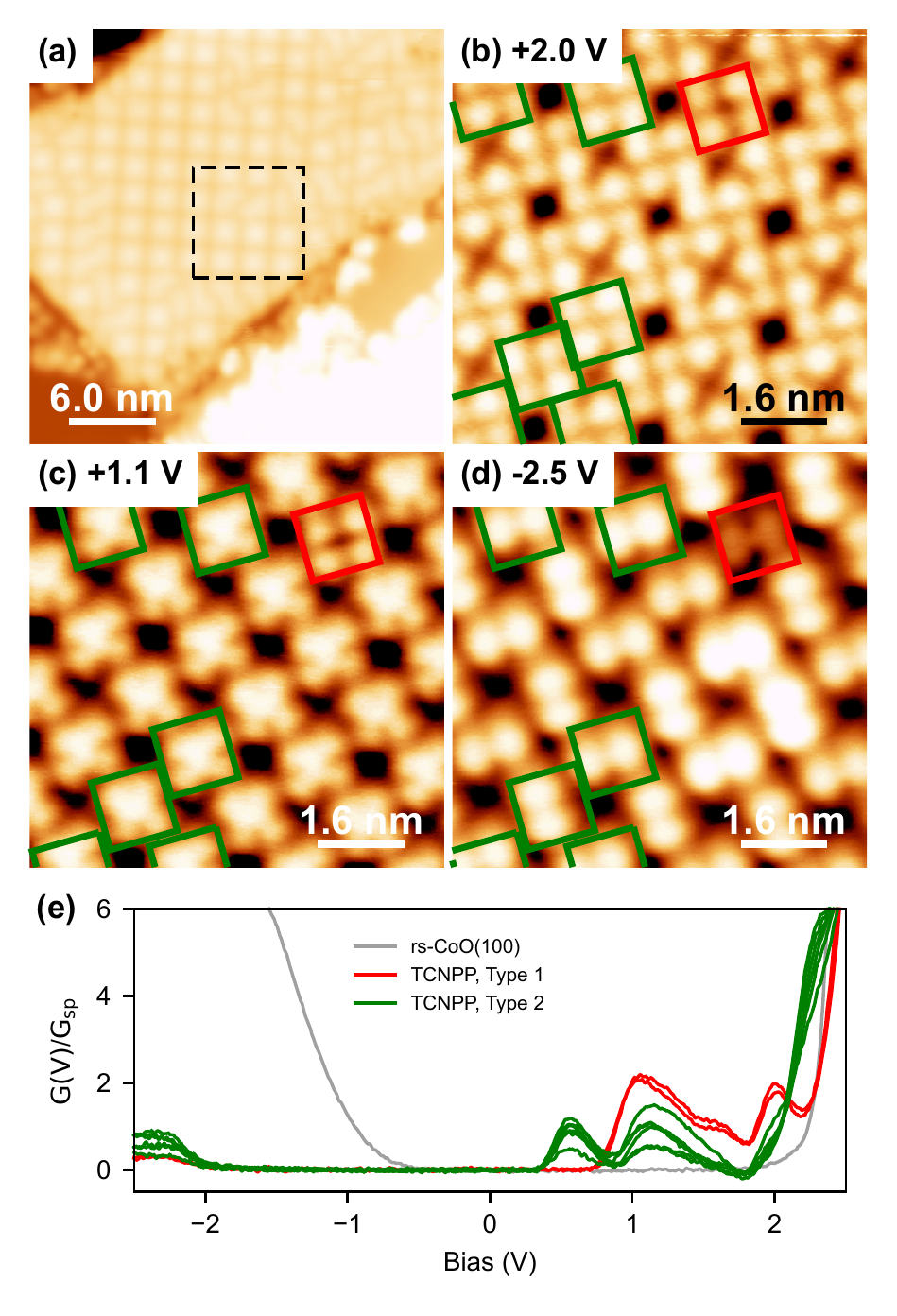}
	\caption{\label{fig:STM_AnSTS}STM/STS measurement of molecules after annealing the sample to 420\,K. (a) Overview of the area. The \rsCoO\ reference is taken in the very lower left corner. (b-c) Bias-dependent topographies similar to \figref{fig:STM_RTSTS}. Almost all molecules in this area are type 2 molecules (green). However, type 1 molecule are also found (red). (e) Normalized conductance STS data taken on the centers of the marked molecules proves the assignment. 
		STM parameters: (a) \stmparameter{3.5}{0.05}; (b) \stmparameter{2.0}{0.05}; (c) \stmparameter{1.1}{0.05}; (d) \stmparameter{-2.5}{0.05}; (e) prior to opening the feed-back loop $G_{sp}:$ \stmparameter{2.5}{0.4}.}
\end{figure}

The analysis shown in \figref{fig:STM_AnSTS} also confirms the assignment with respect to the electronic fingerprint of the molecules. 
From our experiments we conclude that by annealing the \TCNPP\ on \rsCoO\ the dominant molecular species shifts from type 1 to type 2 molecules.  
Therefore we propose that type 2 molecules are in fact Co-TCNPP (see Section \ref{sec:DFT}) the metalation process itself is activated at $T \geq \SI{300}{\K}$. 
By comparison with STM images of Co-tetraphenylporphyrin on Cu(111) \cite{WeberBargioni2008}, the appearance of type 2 molecules in our experiments could point to a relatively flat porphyrin macrocycle.  
The Co atoms most likely come from step edges, grain boundaries of the \rsCoO, or from the defects on some \rsCoO\ terraces.

\subsection{Density Functional Theory\label{sec:DFT}}

The treatment of cobalt oxide by DFT is plagued with aspects that make the analysis of molecular adsorption on that material challenging. 
The correlation effects do not allow to reproduce the semiconducting nature of the material without explicit considerations in the calculations. 
This calls for at least a DFT+U approximation.  
Furthermore, our work on  wurtzite CoO \cite{Ammon2021} has shown that a plain DFT+U treatment would find the wurtzite polymorph energetically more favorable than the rock salt polymorph. 
This deficiency may be healed by inclusion of van der Waals corrections which are also required to account for the adsorption energy of large organic molecules.
The omission of van der Waals corrections leads to arbitrary structural relaxations of the substrate slab with its large lateral unit cell that have nothing to do with the adsorption of the molecule.
The antiferromagnetic coupling between ferromagnetic (111) planes has also to be taken into account to properly obtain a semiconducting \rsCoO\ slab. 
Therefore a spin-polarized calculation must be performed using a correspondingly large unit cell of $(8\times8)$ to obtain electronic properties.

\begin{figure}[t!]
	\centering
	\includegraphics[width=\figWidth]{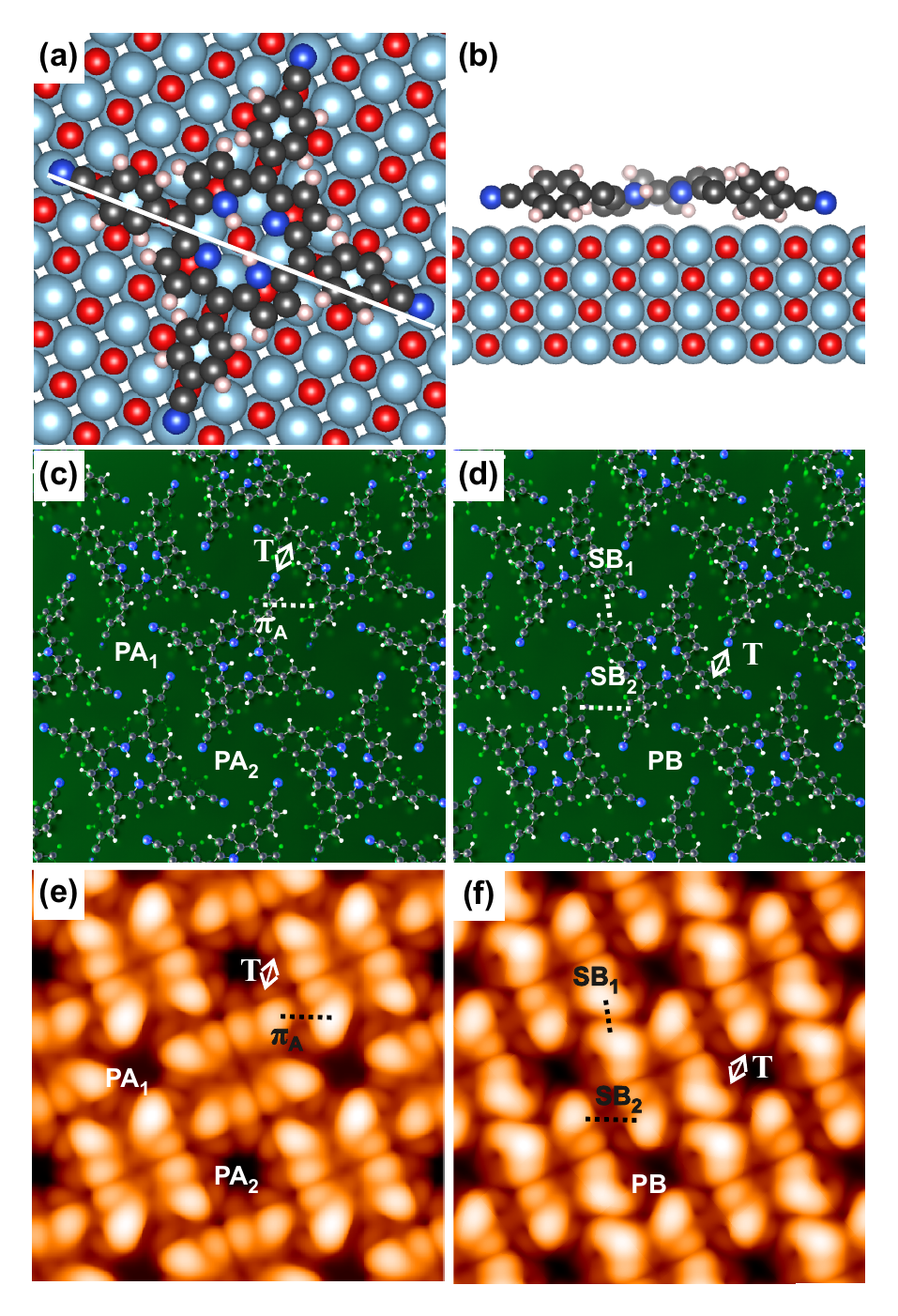}
	\caption{\label{fig:DFT_SingleMol}
		(a) Top view of the energetically most favorable configuration of \TCNPP\ on \rsCoO. The geometry allows for a favorable interaction of the molecular N atoms with the Co ions of the substrate.  (b) Side view, cut along the white line in (a). The molecule has the saddle shape configuration in which the iminic nitrogen atoms are pulled to the substrate. The cyanophenyl groups tend to be pulled parallel to the substrate but steric hindrance does not allow a perfectly flat geometry. The interaction between the CN group and the substrate comes apparent from the bending of the group towards the CoO in (b). 
	(c) and (d): The resulting pore geometries PA$_1$, PA$_2$, and PB in the proposed assemblies A  and B.  A semi-transparent green plane was added to illustrate the orientation of the phenyl rings. Possible intermolecular interactions are labeled by T, $\pi_\textrm{A}$, SB$_1$, and SB$_2$. 
	(e) and (f) corresponding empty-state DFT STM simulation of motif A and B. See main text for details about (c)-(f). }
\end{figure}

In this spirit, we discuss first the adsorption geometry and the occurrence of the two self-assembly motifs found in experiment.
The structural findings of single adsorbed \TCNPP\ are summarized in \figref{fig:DFT_SingleMol}(a) and (b). 
The adsorption site with the center of the molecule sitting on-top of a substrate O ion is energetically most favorable, as is the strict alignment of the molecular axis along the [001] bulk direction of CoO. 
(For the definition of the molecular axes see \figref{fig:MET_schema}.)
The molecule is found in a saddle-shape geometry where macrocycle pyrrole rings turn outward by \SI{17.8}{\degree}, while the deprotonated pyrrole rings turn inwards by \SI{19.3}{\degree} with respect to the surface normal.
We find  the plane of the phenyl rings to be at \SI{36}{\degree} angle with respect to the surface normal.
A slight bending of the terminal -CN group resulting in a calculated N-Co distance of $\SI{3.0(1)}{\AA}$ indicates an attractive interaction with the substrate.

In \figref{fig:DFT_SingleMol}(c, d) we analyze the orientation of the phenyl rings in motifs A and B employing the non-spinresolved DFT calculations in a \rt\ slab (notation with respect to the CoO(100)-($1\times 1$) unit cell). 
In motif A two pore geometries ($\textrm{PA}_1$, $\textrm{PA}_2$), a T-type, and a $\pi$-stacking interaction configuration ($\pi_A$) exist (see \figref{fig:DFT_SingleMol}(c)). 
At $\textrm{PA}_{1}$ all surrounding phenyl rings points upwards like a roof surrounding the pore center. 
At $\textrm{PA}_{2}$ all groups are pointing downwards forming a square cone. 
This allows all phenyl ring pairs of neighboring molecules to be aligned in $\pi$-$\pi$ stacking geometry ($\pi_A$). 
However, the distance between the phenyl rings ($\approx\SI{6}{\angstrom}$) is too large for any bonding relevance.
In the DFT STM simulations the pore $\textrm{PA}_1$ appears as the cross-like pore and $\textrm{PA}_2$ as the square pore (\figref{fig:DFT_SingleMol}(e)). 
The simulation of how the pores are imaged matches nicely with the experimental results (\figref{fig:STM_RTSA}(a, c) and \figref{fig:STM_RTSTS}(a, c)). Concerning the imaging mode of the molecules themselves, a close match is not expected since the non-spinresolved calculations result in a metallic and not a semiconducting CoO substrate. 

In motif B, all pores have the same geometry (PB) with two rings on the opposite sides of the pore pointing upwards and two downwards (\figref{fig:DFT_SingleMol}(d)). 
As a consequence the phenyl rings cannot achieve a $\pi$-stacking configuration as the top edges of phenyl rings of two neighboring molecules point either towards ($\textrm{SB}_1$) or away from each other ($\textrm{SB}_2$). 
The modulation of this phenyl orientation is well visible in the experimental data for \SI{2}{\V} (compare \figref{fig:STM_RTSA}(a) or \figref{fig:STM_RTSTS}(a)). $\textrm{SB}_1$ smears almost out to one single protrusion while at the place of $\textrm{SB}_2$ two separate  protrusions are visible.
The two upwards and downwards pointing groups of pore PB create a symmetry axis which is clearly visible in simulation and experiment as a dark line.
 
In the self-assemblies CN-groups are oriented at $90^\circ$ angle to the axis of the cyanophenyl ring of a neighboring molecule, as observed on other substrates before \cite{Lepper2017,Xiang2020,Stoer2019,Kuliga2021}. 
For the stabilization of the supramolecular assembly via the CN-groups \cite{Yokoyama2001} hydrogen bonding to the pyrrole group \cite{Stoer2019} or the phenyl rings \cite{Lepper2017} has to be considered. 
From our DFT-optimized structure of the assemblies we suggest that the bond to the pyrrole entity (-NH- containing, flat-lying) is dominating, we find an average bond distance of 2.8 and 2.6\,\AA\ for assemblies A and B respectively.
In contrast, the CN...HC distance to a hydrogen of the phenyl ring of a neighboring molecule is at least 3.0\,\AA. This is still in a plausible range for hydrogen binding but the interaction is weaker.

\begin{figure}[t!]
	\centering
	\includegraphics[width=\figWidth]{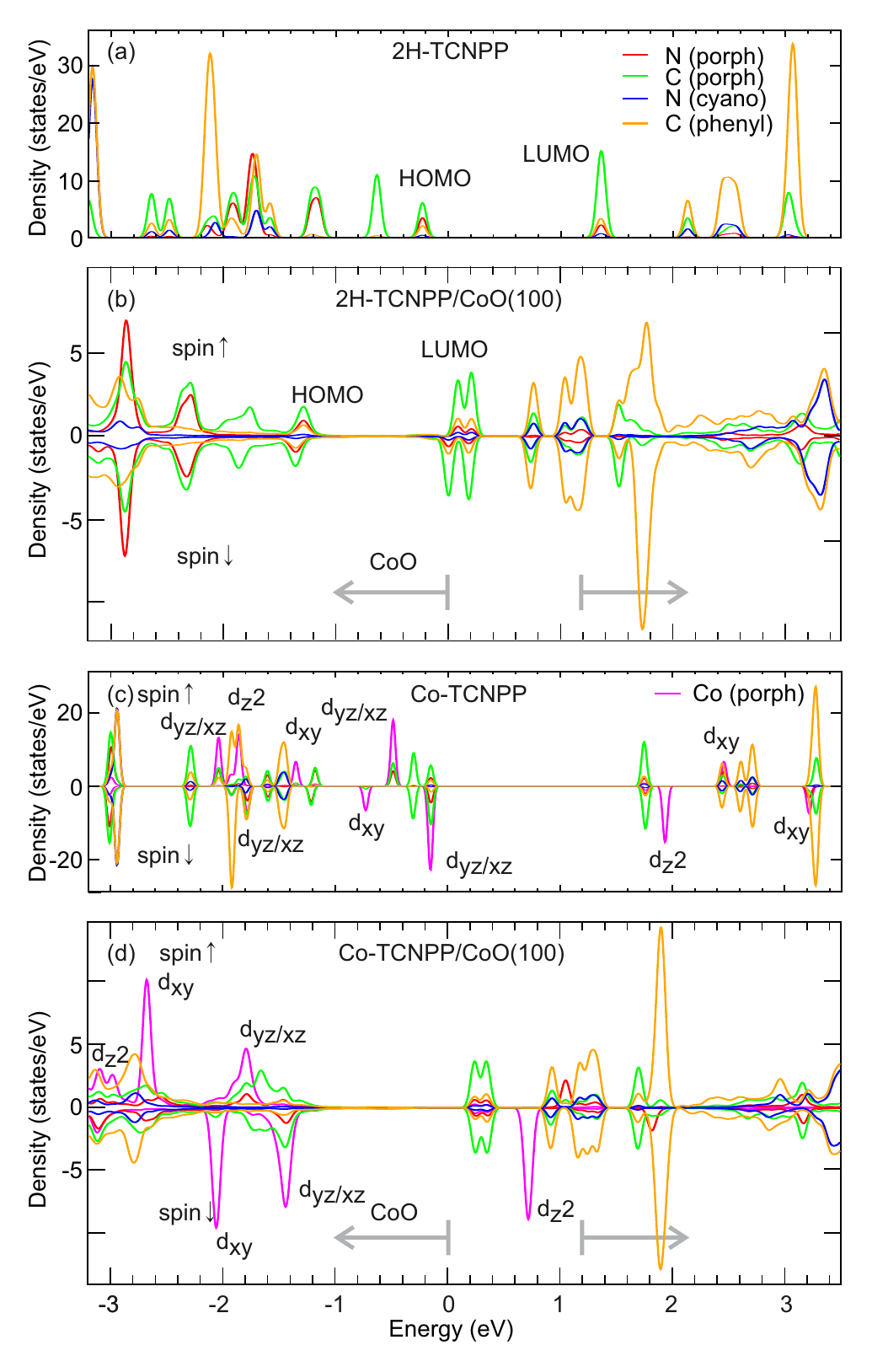}
	\caption{\label{fig:DFT_electronic}Calculated density of states (DOS) projected onto the N and C atoms of the molecular macrocycle (``porph'') and of the phenyl groups (``phenyl'' / ``cyano'') respectively.
	(a) DOS of \TCNPP\ in vacuum. (b) Spin-resolved DOS of \TCNPP\ adsorbed on  \rsCoO. The calculated valence and conduction band regions of \rsCoO\ are indicated by gray arrows.
    (c) Spin-resolved DOS of Co-TCNPP in vacuum and (d) adsorbed on CoO, additionally showing the DOS projected onto the coordinated Co atom of the molecule. Co-DOS is labeled by the orbital character of the d-states. For the Co atom in the molecule $U_\textrm{eff} = 2\,\textrm{eV}$, for the Co in the CoO $U_\textrm{eff} = 4\,\textrm{eV}$ was chosen (see text).
    The x-axis is the same for all panels, its origin is the energy of the highest occupied state of the respective calculation.
  	}
	
\end{figure}

In \figref{fig:DFT_electronic} we summarize the calculated electronic properties of \TCNPP\ and Co-TCNPP in vacuum and on \rsCoO\ using spin-polarized PBE+U+TS. 
This procedure maintained the semiconducting nature of CoO with a calculated band gap of \SI{1.2}{eV}.
We project the DOS to characteristic groups of atoms of the molecule which allows us to follow the shift of electronic states upon adsorption.
We notice that the HOMO-LUMO gap of \TCNPP\ in vacuum appears between states dominantly located at the porphyrin core (\figref{fig:DFT_electronic}(a)). 
The characteristic pattern of HOMO, \mbox{HOMO-1} and LUMO, LUMO+1 (and following states) can also be recognized in case of \TCNPP\ on \rsCoO\ (\figref{fig:DFT_electronic}(b)). 
The LUMO orbitals shift down in energy, LUMO(+0) gets slightly filled whereas LUMO+1 and LUMO+2 remain in the gap at the conduction band edge of CoO.
The HOMO states also shift down, the HOMO-LUMO separation remains approximately the same as for the free molecule.

The DOS of the Co-TCNPP was calculated using $U_\textrm{eff} = \SI{2}{eV}$ vs. $U_\textrm{eff} = \SI{4}{eV}$ for the CoO as motivated by \cite{Leung2010}. 
The consequence of a smaller $U_\textrm{eff}$ value is that the Co d-states move closer to the Fermi level. 
Obviously, CoO bandgap and molecular Co d-level positions depend sensitively on the choice of the DFT approximation which limits the meaningfulness of the comparison to experimental findings somewhat.
Nonetheless, we propose the following conclusions. 
The calculations do not indicate significant electron transfer between molecule and substrate, hence the experimental $\alpha$- and $\beta$-peaks in the band gap should correspond to the lowest lying LUMO states. 
The main difference between experimental spectra of \TCNPP\ and Co-TCNPP adsorbed on CoO in the energy ranged probed can be explained by the appearance of an unoccupied, $\textrm{d}_{\textrm{z}^2}$ derived Co state, which is also responsible for the ``filled'' molecular center in the STM images.

The calculations also support the view that the HOMO orbital should be located well below the valence band onset.
The experimental spectra of \TCNPP\ on CoO(100) of \figref{fig:STM_RTSTS}(e) seem to suggest the absence of any molecular states in an energy range exceeding 3\,eV which seems too large to be identified with the HOMO-LUMO gap of the adsorbed molecule. 
We therefore propose that the $\alpha$-peak in \figref{fig:STM_RTSTS}(e) should not be identified with a LUMO+0 state. 
We think it is composed of the states LUMO+1 to LUMO+3 and $\beta$ partially of LUMO+3 and of following states, as LUMO+3 is most likely already affected by hybridization with states at the conduction band minimum.  
Using the spin-resolved electron density to calculate a spin-averaged STM image of \TCNPP, we notice that the clear characteristic ``four-bump appearance'' observed in experiment at a bias voltage of 1.1\,V (i.e. beyond the $\alpha$-peak) only starts to appear after charge density integration beyond the LUMO+2 slightly below 1.4\,eV but is more clearly developed when the LUMO+3 is included, the corresponding STM simulation is shown in \figref{fig:STM_RTSA}(c). 
This indicates that $\alpha$ contains at least contributions from LUMO+1 and LUMO+2.

However, if the LUMO+0 state were to be included into the $\alpha$-peak, it would have to be strongly modified by hybridization with the substrate (within the fundamental gap of CoO) to reduce the energetic distance to the LUMO+1 and following states for which there is no apparent reason. 
Similarly the appearance of the unoccupied Co $\textrm{d}_{\textrm{z}^2}$ state $\gamma$ energetically just  below the state $\alpha$ (\figref{fig:STM_RTSTS}(e)), is more consistent with a scenario where $\alpha$ does not contain the LUMO+0 state.
Experimentally, a state expected at or near the Fermi energy ($V_\textrm{bias}=\SI{0}{\V}$) within the fundamental gap of an (undoped) semiconductor may not be available as transport channel due to its midgap position and hence can potentially not be detected in a tunneling experiment.	
The clarification of this needs a much more rigorous theoretical treatment of this complex system.

\section{Conclusions}

By careful tuning of preparation parameters we demonstrated the growth of \rsCoO\ films on Au(111) suitable for adsorption studies. 
Films prepared with thicknesses of three to five CoO layers have the advantage that their electronic properties are close to bulk-like with a band gap determined by STS of $E_\mathrm{g}=\SI{2.5(2)}{\eV}$.
In that thickness range clean Au(111) areas are left that serve as a reference for calibrating STM tip properties.
The \rsCoO\ films may serve as a more reactive, easier to prepare, and structurally more perfect alternative to MgO \cite{Lin2009,Pal2014,Hurdax2020} or NaCl films to decouple molecules from a metal surface \cite{Repp2005,Liljeroth2007, Wang2015a}. 
Furthermore, due to the smaller band gap even thicker, homogeneous bulk-like oxide films can be prepared, suitable for a large range of surface-science techniques.

As an example, the adsorption, self-assembly, self-metalation and electronic properties of sub-monolayer \TCNPP\ on \rsCoO\ were studied. 
By a combination of STM and DFT analysis, we find that the molecules adsorb flat-lying, in a saddle-shape conformation, with their nitrogen atoms interacting weakly with underlying Co ions. 
When deposited at 300\,K the \TCNPP\ molecules assemble into two different, and hence energetically close or degenerate superstructures. 
The appearance of the superstructure in STM agrees well with DFT simulated images.
From DFT structural relaxations we suggest that the self-assembly is stabilized by hydrogen bridge bonds between the nitrile terminal goups of \TCNPP\ and hydrogen atoms from the molecular porphyrin core. 
By STS different configurations of the molecules are identified. 
One of them which is the majority species when the molecules are adsorbed at 300\,K is assigned to \TCNPP. 
A second configuration becomes dominant when annealing the system to 420\,K. 
From the appearance of the molecules in STM, we propose that a self-metalation reaction has taken place and that the second species is \mbox{Co-TCNPP}.
DFT calculations of the electronic density of states support this view and tentatively allow to identify the features of the STS measurements. 
It is found that upon adsorption the molecular states shift down in energy to an extent that the HOMO-LUMO gap of the free molecule is shifted into the valence band region of CoO, effectively suppressing the tunneling conductivity in that energy range.  

\section{Acknowledgements}
This work was supported by the Deutsche Forschungsgemeinschaft (DFG) within the research unit ``Functional Molecular Structures on Complex Oxide Surfaces (\textit{fun}COS)'' (project no. 214951840). M.A.S. gratefully acknowledges support and supply of CPU time by the Regionales Rechenzentrum Erlangen (RRZE).  

 \bibliographystyle{apsrev4-2}
 \bibliography{AdsorptionSelfassemblyTCNPPonCoO100_arXiv}
\end{document}